\documentclass[10 pt,conference, letterpaper]{IEEEtran}

\usepackage[dvips]{graphicx}

\usepackage{subfig}
\usepackage[cmex10]{amsmath}
\usepackage{amsfonts}
\usepackage{amssymb}
\usepackage{times}
\usepackage{citesort}
\usepackage[usenames]{color}
\usepackage{verbatim}
\usepackage{array}
\usepackage{cite}
\usepackage{ellipsis}
\usepackage{url}
\usepackage[]{algorithm}
\usepackage{algorithmic}
\usepackage[T1]{fontenc}
\usepackage{amsmath}
\usepackage{multirow}
\usepackage{tabularx}
\usepackage{caption}

\begin{document}
\graphicspath{{figures/}}
%
\title{Technical Rate of Substitution of Spectrum in Future Mobile Broadband Provisioning}

\author{Yanpeng~Yang~and~Ki Won Sung
\\
  KTH Royal Institute of Technology, Wireless@KTH, Stockholm, Sweden\\
  E-mail: yanpeng@kth.se, sungkw@kth.se
}
\maketitle

\begin{abstract}
Dense deployment of base stations (BSs) and multi-antenna techniques are considered key enablers for future mobile networks. Meanwhile, spectrum sharing techniques and utilization of higher frequency bands make more bandwidth available. An important question for future system design is which element is more effective than others. In this paper, we introduce the concept of technical rate of substitution (TRS) from microeconomics and study the TRS of spectrum in terms of BS density and antenna number per BS. Numerical results show that TRS becomes higher with increasing user data rate requirement, suggesting that spectrum is the most effective means of provisioning extremely fast mobile broadband.


\end{abstract}

\begin{IEEEkeywords}
Spectum, densification, multi-antenna, stochastic geometry
\end{IEEEkeywords}


%
\IEEEpeerreviewmaketitle

\section{Introduction}

Mobile wireless communication has experienced explosive growth during the last decades. According to \cite{Cisco}, global mobile data traffic grew 69 percent and almost half a billion (497 million) mobile devices and connections were added in 2014. In the near future, global mobile data traffic will increase nearly tenfold between 2014 and 2019, which will challenge the upcoming 5G network. The traffic explosion concerns the enhancement of the user experience and sustainment of the increased traffic volumes, addressed by the two challenges: very high data rate, and very dense crowds of users \cite{METIS}.

There is a general consensus that three fundamental ingredients are available for increasing wireless networks capacity: more spectrum, denser base station (BS) deployment, and better transmission technology with multi-antennas \cite{RealWireless}\cite{NSN}. Small cell deployment has drawn great interest to academia and industry recently \cite{hyper}. The interference mitigation and energy efficiency problems are specifically studied in \cite{Huang, Zhang, Li, Soh, Razavi}. Hence, dense deployment could guarantee high spectral and energy efficiency which qualifies for future. Meanwhile, multi-antenna or even massive multi-input multi-output (MIMO) system is considered as another promising solution. By adding multiple antennas, it offers a spatial dimensional in addition to time and frequency for communications and yields a degree of freedom gain to accommodate more data. Therefore, a significant performance improvement can be obtained with regard to reliability, spectral efficiency, and energy efficiency \cite{Rusek}. Different from the other two elements implemented in infrastructure, spectrum is considered as a virtual resource. In Shannon's model, the additional bandwidth can lead to a linear increase in data capacity. However, spectrum is rather scarce due to cost and authorization reasons. Some efforts have been made to better utilize spectrum, e.g. licensed shared access (LSA) and authorized shared access (ASA)\cite{Ahokangas}. Also, the feasibility of millimeter-Wave band has been widely tested for future wireless systems \cite{ZPi, Rappaport, Roh}.

Given the three factors, design of future mobile systems involves a decision on which direction to head to, in order to meet the enormous traffic demand efficiently. It is shown that excessive deployment leads to trivial effect in some circumstances \cite{Yang}. On the other hand, considering the rarity of spectrum, when and where spectrum is needed most remains to be answered. Thus, it is necessary to know about the relationship among the three ingredients for various scenarios, which lacks thorough investigation. In \cite{Yang}, the authors have shown the tradeoff between spectrum and densification with a certain technology. BS densification performs well in sparse network where user throughput increases almost linearly with the BS density. On the contrary, further densification is not effective in dense networks. However, multi-antenna techniques and other types of traffic demand such as usage increase are not considered in that work. Usage increase can be interpreted as either the data consumption of each user rises or active user density grows, e.g., in stadium and metropolitan area. Compared with high data rate demand situation, the resource requirement for high usage is not fully studied.


In this paper, we introduce microeconomic terminology 'technical rate of substitution' (TRS) \cite{Varian} to illustrate the value of spectrum in terms of BS density and antenna number per BS. TRS measures the tradeoff between two inputs resulting in same production. Stochastic geometry based framework is applied for analysis. The substitutability of spectrum is discussed by calculating the TRS between spectrum and other elements in different scenarios. Furthermore, we investigate how resource requirement changes when upgrading a network for two purposes: increasing the individual user data rate or accommodating more active users, which constitute the same area capacity increase. Our study can give insights into the proper choice of allocating resources for specific purposes in different scenarios.

The remainder of this paper is organized as follows: Section \ref{sec:SM} describes our system model and problem formulation. Then, Section \ref{sec:Methodology} introduces our methodology. In Section \ref{sec:Result}, we present and analyze the numerical results. Finally, the conclusions are discussed in Section \ref{sec:Con}.

\section{System Model and Problem Formulation}
\label{sec:SM}

\begin{figure}[t]
\includegraphics[width=.5\textwidth]{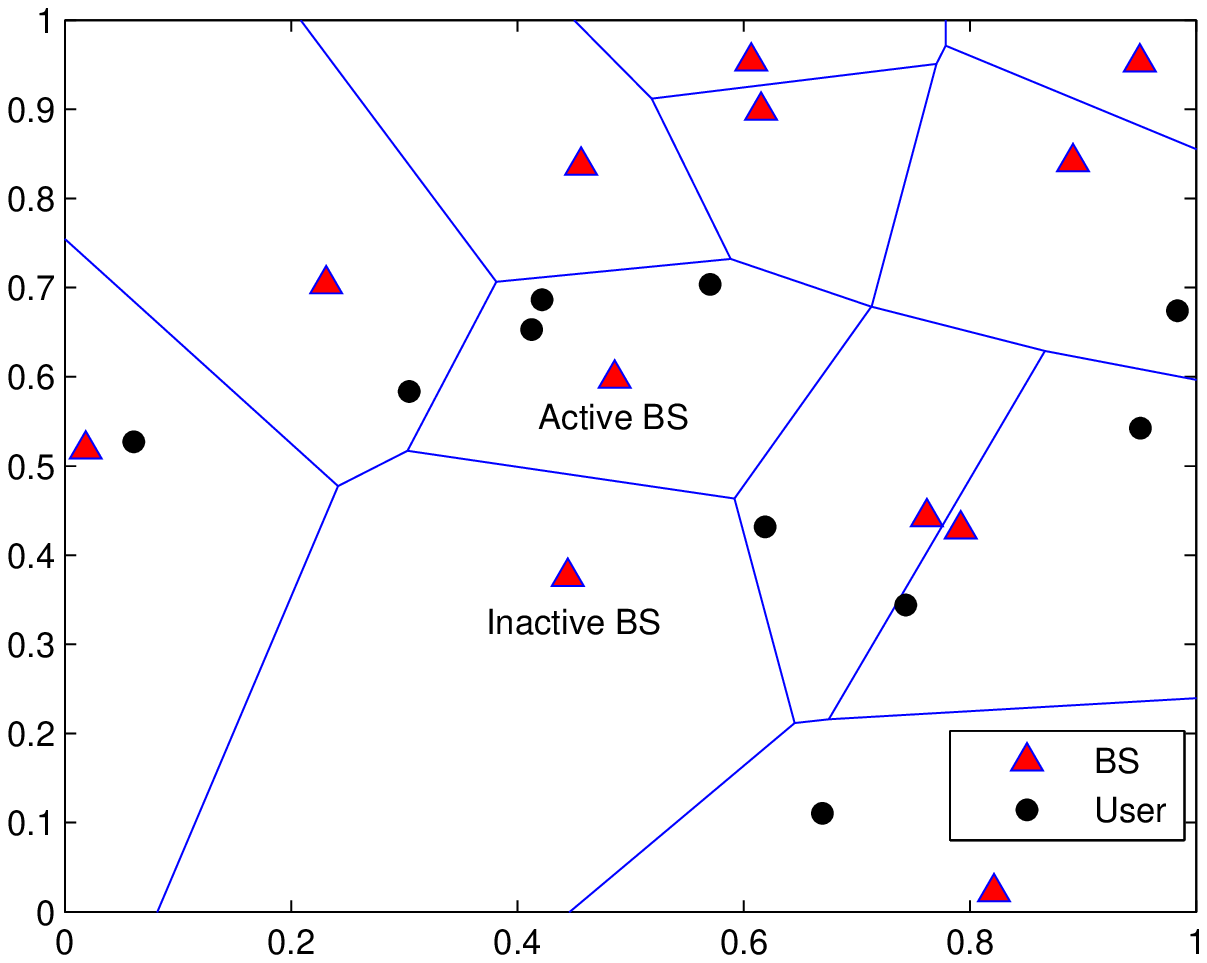}
\caption{Network Layout.}
\label{fig:Voronoi}
\end{figure}
We consider the downlink transmission of a wireless network. On the supply side, BSs are deployed in the network with density $\lambda_{b}$, each of which is equipped with M antennas. The amount of bandwidth allocated to the network is \emph{W} and universal frequency reuse scheme is applied. On the demand side, the density of active users is $\lambda_{u}$ and each user device has a single antenna. We define basic service as providing data rate $R_{a}$ to the active users in the network. The basic service can be attained with different resource combinations, e.g., installing more antennas to BSs meanwhile reducing the total spectrum and etc. Thus, which resource is more effective and efficient is a crucial question to the system design and deployment. In our work, we will compare the effect of spectrum and infrastructure (BS and antenna) in network dimensioning.

A concept from microeconomics, technical rate of substitution, is introduced to explain the value of spectrum. TRS measures the rate at which the firm will have to substitute one input for another in order to keep output constant. Suppose that we have factors \textit{1} \& \textit{2} and are operating at some point ($x_{1},x_{2}$) with output $y$. Consider a change in $x_{1}$ and $x_{2}$ while $y$ keeps fixed:
\begin{equation}
\label{eqn:output}
y=f(x_{1},x_{2})=f(x_{1}+\triangle x_{1},x_{2}+\triangle x_{2}).
\end{equation}

Then the TRS of factor \textit{1} with regard to factor \textit{2} at point ($x_{1},x_{2}$) is defined as
\begin{equation}
\label{eqn:TRS}
\mathbf{TRS}(x_{1},x_{2})=\left|\frac{\triangle x_{2}}{\triangle x_{1}}\right|
\end{equation}
which indicates the substitutability of factor \textit{1}. If the TRS of factor \textit{1} is low, it means factor \textit{1} can be replaced by factor \textit{2}. Otherwise, factor \textit{1} overweighs factor \textit{2}.

Our objective is to calculate $\mathbf{TRS}(W,\lambda_{b})$ and $\mathbf{TRS}(W,M)$ at each operating point of basic service. Beyond that, we consider increasing the network capacity from either speed or usage perspective and investigate the resource requirement for the two network upgrading purposes.

\section{Methodology}
\label{sec:Methodology}
In this work, stochastic geometry is used to model the network. The BSs and users follow two independent homogeneous poisson point process (\textbf{PPP}) $\Phi_{b}$ and $\Phi_{u}$ in $\mathbb{R}^{2}$ with intensities $\lambda_{b}$ and $\lambda_{u}$, respectively. We assume each user is associated with the closest BS. Namely the coverage area of each BS comprises a Voronoi tessellation, which is illustrated in Fig. \ref{fig:Voronoi}. Due to the independency between BSs and users, there may be some BSs have no users in their Voronoi cells at some given time slots. Those BSs in the empty cells are called inactive BSs and they are considered not transmitting any signals at this time. When there are more than one user in a cell, the BS serves the users with Round-robin scheduler. We denote the active probability of a typical BS in the network as \emph{$p_{a}$} and BS-user density ratio as $\rho\triangleq\frac{\lambda_{b}}{\lambda_{u}}$. In \cite{Kim}, \emph{$p_{a}$} can be expressed as a function of $\rho$
\begin{equation}
\label{eqn:Pa}
\emph{$p_{a}$} = 1-(1+\frac{1}{3.5\rho})^{-3.5}.
\end{equation}

Maximal ratio transmission (MRT) beamforming is chosen for multi-antenna technique and the standard power loss propagation model is considered with path loss exponent $\alpha$>2. For a typical user served by BS \emph{i}, the average signal to interference plus noise ratio (SINR) is defined as
\begin{equation}
\label{eqn:SINR}
\textrm{SINR} = \frac{g_{ii}P_{t}r_{ii}^{\alpha}}{\sum_{j\in\hat{\Phi}_{b}\textbackslash i} g_{ji}P_{t} + \sigma^{2}},
\end{equation}
where $P_{t}$ is the transmit power, $\sigma^{2}$ denotes the additive noise power. $r_{ji}$ and $g_{ji}$ are the distance and channel gain from BS $\emph{j}$ to the user in cell $\emph{i}$ and $\hat{\Phi}_{b}$ represents the set of active BSs. According to \cite{Lee}, $\sum_{j\in\hat{\Phi}_{b}}$ is assumed to be a homogeneous \textbf{PPP} with intensity $p_{a}\lambda_{b}$ for simplifying the analysis.
The outage probability is given by \cite{Li2}
\begin{equation}
\begin{split}
\begin{aligned}
\label{eqn:Pout}
p_{out} &= \mathbb{P}[\textrm{SINR}<T]\\
&= 1-\frac{1}{\emph{$p_{a}$}}\left\Vert \left[\left(k_{0}+\frac{1}{p_{a}}\right)\textbf{I}-\textbf{Q}_{M}\right]^{-1} \right\Vert_{1},
\end{aligned}
\end{split}
\end{equation}
where $\|\cdot\|_{1}$ is the $\emph{L}_{1}$ induced matrix norm, \textbf{I} is an $M \times M$ identity matrix. $\textbf{Q}_{M}$ is an $M \times M$ Toeplitz matrix given by
\begin{equation}
\label{eqn:QM}
\textbf{Q}_{M} = \left(
                   \begin{array}{ccccc}
                     0 &   &   &   &   \\
                     k_{1} & 0 &   &   &   \\
                     k_{2} & k_{1} & 0 &   &   \\
                     \vdots &  &  & \ddots &   \\
                     k_{M-1} & k_{M-2} & \cdots & k_{1} & 0 \\
                   \end{array}
                 \right),
\end{equation}
$k_{0}=\frac{\frac{2}{\alpha}T}{1-\frac{2}{\alpha}}\,_2F_1(1,1-\frac{2}{\alpha};2-\frac{2}{\alpha},-T)$, and $k_{i}=\frac{\frac{2}{\alpha}T^{i}}{i-\frac{2}{\alpha}}\,_2F_1(i+1,i-\frac{2}{\alpha};i+1-\frac{2}{\alpha},-T)$ for $i \geq 1$, where $\,_2F_1(\cdot)$ is the Gauss hypergeometric function.

The average achievable user data rate $R_{a}$ can be expressed as a function of $p_{out}$ and is given by
\begin{equation}
\begin{split}
\begin{aligned}
\label{eqn:Ra}
R_{a} &= W\mathbb{E}[\textrm{log}_{2}(1+\textrm{SINR})]\\
&= \frac{W}{\textrm{ln}2}\int_{t>0}\mathbb{P}(\textrm{ln}(1+\textrm{SINR})>t)dt\\
&= \frac{W}{\textrm{ln}2}\int_{T>0}\frac{\mathbb{P}(\textrm{SINR}>T)}{T+1}dT\\
&= \frac{W}{\textrm{ln}2}\int_{T>0}\frac{1-p_{out}}{T+1}dT\\.
\end{aligned}
\end{split}
\end{equation}


\section{Numerical Results}

\label{sec:Result}
Two scenarios are considered in this paper: sparse network representing current deployment and dense network for the future. Sparse and dense are interpreted by the relative BS-user density ratio $\rho$. We choose $\rho=0.1$ and $\rho=10$ for sparse and dense scenario respectively while $\lambda_{u}$ is fixed to 100 per area. We set path loss exponent $\alpha$ to 4 for simplification and ignored the additive noise which is proved to be negligible in \cite{Li2}\cite{Andrews}. The resource requirement for a twofold increase in data rate or usage is studied as capacity expansion.

\begin{figure}[t!]
\centering
\subfloat[Sparse Network]{\label{fig:AntSpSparse}\includegraphics[width=0.5\textwidth]{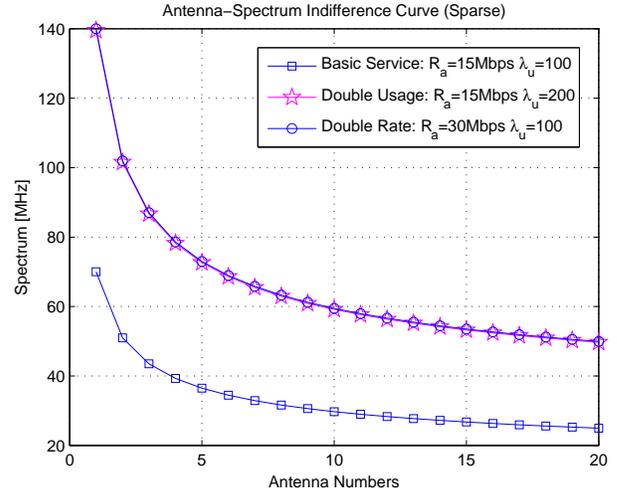}}\\
\subfloat[Dense Network]{\label{fig:AntSpDense}\includegraphics[width=0.5\textwidth]{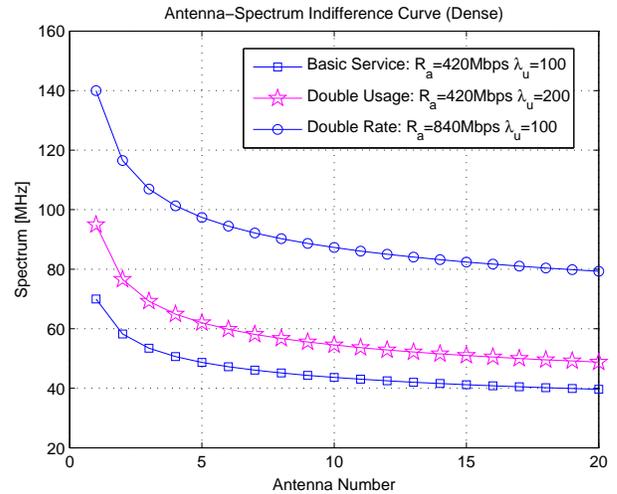}}
\caption{Relation between spectrum and antenna number.}
\label{fig:AntSp}
\end{figure}

The relationship between spectrum and antenna number in both scenarios are illustrated in Fig. \ref{fig:AntSp}. The indifference curves reflect resource combinations for the same traffic demand. For basic service, adding multi-antennas has a diminishing gain in both scenarios. This is reflected in the growing TRS between spectrum and antenna number shown in Table \ref{table:TRS1}. In Fig. \ref{fig:AntSpSparse}, the coincident indifference curves indicate that doubling usage requires equal amount of resource with doubling data rate in sparse network. Since user SINR distribution does not change when more active users appear in this scenario, double spectrum can serve $\lambda_{u}$ with $2R_{a}$ or $2\lambda_{u}$ with $R_{a}$. Nevertheless, this does not apply to the dense regime where double usage will change the user SINR distribution. In dense deployment, doubling usage requires less resources because the inactive BSs can turn on and complement the effect of spectrum. From the antenna perspective, increasing to 6 antennas per BS can double the data rate (usage) in sparse case. In dense scenario, installing 3 antennas will double the usage of single antenna case. However, it is almost impossible to double the data rate by adding antennas.

\begin{figure}[t!]
\centering
\subfloat[Sparse Network]{\label{fig:SpDenSparse}\includegraphics[width=0.5\textwidth]{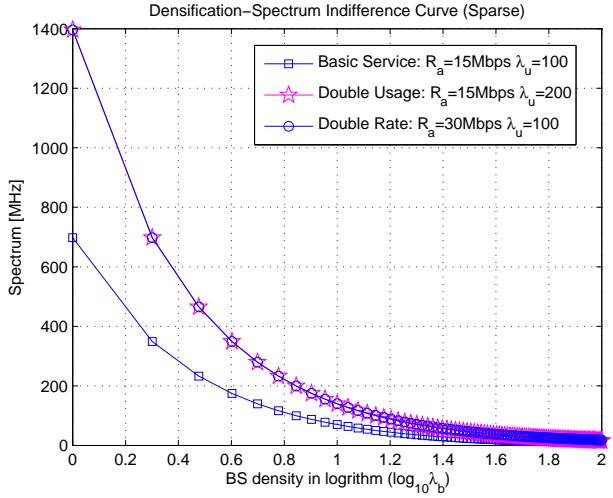}}\\
\subfloat[Dense Network]{\label{fig:SpDenDense}\includegraphics[width=0.5\textwidth]{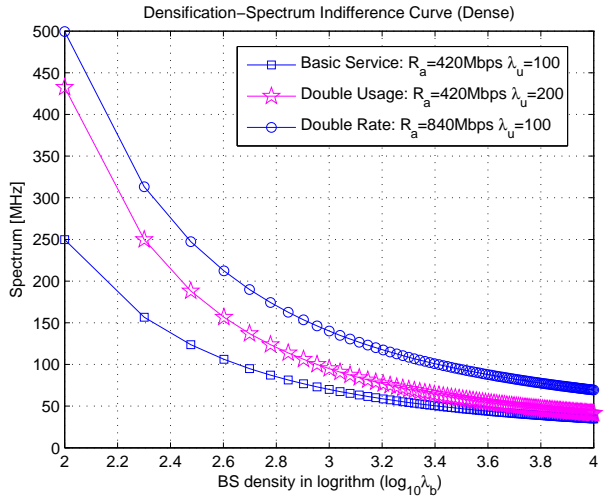}}
\caption{Relation between spectrum and BS density.}
\label{fig:SpDen}
\end{figure}

Figure \ref{fig:SpDen} depicts the spectrum-densification relation in both scenarios. In sparse network (Fig. \ref{fig:SpDenSparse}), similar with the discussion above, same amount of resources are needed for doubling data rate and usage. Besides, in this situation, TRS between spectrum and densification is rather low as shown in Table \ref{table:TRS2}, which means densification is more effective. In dense scenario (Fig. \ref{fig:SpDenDense}), twofold densification can still achieve double usage but is gradually getting farther from reaching double data rate. This phenomenon indicates that doubling data rate in already dense region requires significant amount of resources in terms of infrastructure. Contrarily, the required amount of spectrum is relatively small, which implies that the TRS is extremely high for high data rate dense scenario. Thus, acquiring a few more spectrum may have a huge impact and save the required BS deployment.

In general, infrastructure improvement can manage the traffic demand regarding both volume and speed in current macro networks. The low TRS indicates that spectrum is replaceable in sparse networks. The value of spectrum is reflected better in future high data rate densely deployed networks. In this case, spectrum seems to be the only solution when technology and deployment are hardly effective.

\begin{table}[h!]
\captionsetup{font=scriptsize}
\caption{TRS between Spectrum and Antenna number ($\frac{\triangle M}{\triangle \textrm{MHz}}$)}
\label{table:TRS1}
\centering
\begin{tabular}{|>{\centering}m{1.5cm}|>{\centering}m{1cm}|>{\centering}m{1cm}|>{\centering}m{1cm}|m{1cm}<{\centering}|}
\hline
Sparse Case  &   M=1    &  M=4 & M=8 & M=16     \\\hline
$\lambda_{u}$=100 $\lambda_{b}$=10	& \multirow{2}{*}{0.053} &	\multirow{2}{*}{0.357}	& \multirow{2}{*}{0.990} & \multirow{2}{*}{2.632} \\
$R_{a}$=15Mbps & \multirow{2}{*}{} &	\multirow{2}{*}{} & \multirow{2}{*}{} & \multirow{2}{*}{} \\\hline
Dense Case  &   M=1    &  M=4 & M=8 & M=16     \\\hline
$\lambda_{u}$=100 $\lambda_{b}$=1000	& \multirow{2}{*}{0.083} &	\multirow{2}{*}{0.513}	& \multirow{2}{*}{1.250} & \multirow{2}{*}{3.030} \\
$R_{a}$=420Mbps & \multirow{2}{*}{} &	\multirow{2}{*}{} & \multirow{2}{*}{} & \multirow{2}{*}{}\\
\hline
\end{tabular}
\end{table}

\begin{table}[h]
\captionsetup{font=scriptsize}
\caption{TRS between Spectrum and Densification ($\frac{\triangle\lambda_{b}}{\triangle \textrm{MHz}}$)}
\label{table:TRS2}
\centering
\begin{tabular}{|>{\centering}m{2cm}|>{\centering}m{1.5cm}|>{\centering}m{1.5cm}|m{1.5cm}<{\centering}|}
\hline
 Sparse Case  &   $\lambda_{b}$=1    &  $\lambda_{b}$=5 & $\lambda_{b}$=10     \\\hline
$\lambda_{u}$=100	& \multirow{2}{*}{0.00286} &	\multirow{2}{*}{0.043}	& \multirow{2}{*}{0.159} \\
$R_{a}$=15Mbps & \multirow{2}{*}{} &	\multirow{2}{*}{} & \multirow{2}{*}{} \\\hline
 Dense Case  &   $\lambda_{b}$=100    &  $\lambda_{b}$=500 & $\lambda_{b}$=1000     \\\hline
$\lambda_{u}$=100	& \multirow{2}{*}{0.417} &	\multirow{2}{*}{12.50}	& \multirow{2}{*}{33.30} \\
$R_{a}$=420Mbps & \multirow{2}{*}{} &	\multirow{2}{*}{} & \multirow{2}{*}{} \\
\hline
\end{tabular}
\end{table}

\section{Conclusion and Future Work}
\label{sec:Con}
In this paper, we investigated the value of spectrum for future mobile systems. For this, we measured relative effectiveness of additional spectrum over densification of network infrastructure (BS and multi-antenna) by borrowing the concept of TRS from microeconomics. Numerical results show that the TRS of spectrum varies substantially with user data rate requirement. For serving moderate data rate, TRS of spectrum is low, indicating that the infrastructure can easily substitute for spectrum. On the contrary, additional spectrum is much more effective than densifying BSs or installing more antennas for achieving very high data rate. We also observe that increasing data rate of individual users requires more resources than accommodating more usage with the same rate. Our analysis suggests that spectrum is an indispensable resource for future mobile systems particularly when the very high data rate for individual users is a driving force of the future systems.


This study was based on the assumption of MRT beamforming without network coordination. Future work should consider other multi-antenna techniques and coordination between BSs for comparison. Besides, economic analysis including cost models will give us further insights into the future system design and deployment principles.







%


\end{document}